\documentclass[12pt,preprint]{aastex}

\slugcomment{for submission to Astrophysical Journal Supplement}

\shorttitle{Draco catalog}
\shortauthors{Rave et al.}

\begin{document}

\title{SDSS catalog of stars in the Draco dwarf spheroidal galaxy}

\author{
Heather A. Rave\altaffilmark{\ref{RPI}},
Chongshan Zhao\altaffilmark{\ref{RPI}},
Heidi Jo Newberg\altaffilmark{\ref{RPI}},
Brian Yanny\altaffilmark{\ref{FNAL}},
Donald P. Schneider\altaffilmark{\ref{PSU}},
J. Brinkmann\altaffilmark{\ref{APO}},
Don Q. Lamb\altaffilmark{\ref{UChicago}}
}

\altaffiltext{1}{Dept. of Physics, Applied Physics and Astronomy, Rensselaer
Polytechnic Institute Troy, NY 12180\label{RPI}}

\altaffiltext{2}{Fermi National Accelerator Laboratory, P.O. Box 500, Batavia,
IL 60510\label{FNAL}}

\altaffiltext{3}{Department of Astronomy and Astrophysics, Pennsylvania State University, 
525 Davey Laboratory, University Park, PA 16802\label{PSU}}

\altaffiltext{4}{Apache Point Observatory, P. O. Box 59, Sunspot, NM 88349-0059\label{APO}}

\altaffiltext{5}{Dept. of Astronomy and Astrophysics, University of Chicago, 5640 S. Ellis Ave., Chicago, IL 60637\label{UChicago}}

\begin{abstract}

The Sloan Digital Sky Survey (SDSS) has scanned the entire region
containing the Draco dwarf spheroidal galaxy to $23^{rd}$ magnitude in
$g^*$.  We present a catalog of stars found in a 453 square arcminute, 
elliptical region centered on the Draco dwarf spheroidal galaxy.  
Objects in the catalog are matched with five previously published
catalogs.  The catalog contains SDSS photometry for 5634 individual objects,
and also the photometry from matches to any of the other catalogs.  A 
comparison of the photometry between catalogs allows
us to identify 142 candidate variable objects.  One hundred and
twelve of the suspected variables have colors consistent with RR
Lyrae variables.

\end{abstract}

\keywords{catalogs --- galaxies: dwarf --- galaxies: individual (Draco dwarf spheroidal) --- stars: variable -- stars: RR Lyrae}

\section{INTRODUCTION\label{intro}}

The Draco dwarf spheroidal companion to the Milky Way, at a distance
of about 82 kpc \citep{m98}, was discovered fifty years ago from
examination of Palomar Observatory Sky Survey Schmidt plates \citep{w55}.
The horizontal branch for this dwarf galaxy is located at $g^*
\sim 20$ and it covers more than 1/3 of a square degree on the sky.
Draco is one of the nearest of the Galaxy's known dwarf spheroidal 
companions and is among the faintest known galaxies, with a luminosity of 
$2 \times 10^5 L_\odot$ \citep{gmh98}.  

Dwarf spheroidal galaxies typically contain many
variable stars, predominately the RR Lyrae variables typical of older, metal-poor
stellar populations \citep{s79}.  Draco is no exception.  The first extensive 
study of Draco by \cite{bs61} found 138 variable stars in the central part of Draco.  
They determined that all but five of these variables were RR Lyraes.  
The five remaining variables were deemed to be ``anomalous" Cepheids.  These 
Cepheids have a shorter period and so do not obey the period-luminosity relation 
for type II Cepheids \citep{cs86}.  These ``anomalous" Cepheids 
have been found in other dwarf spheroidals since their discovery in Draco 
\citep{z78}.  Five carbon stars have also been found in Draco by 
\citet{als82} and \citet{metal2002}.
More recent compilations of variables in Draco are underway \citet{k02}.

The first detailed photometric survey of the Draco dwarf spheroidal galaxy 
was created by \cite{bs61}.  Since then Draco has become one of 
the best studied dwarf galaxies with several more recent photometric 
surveys.  We have created our own catalog of stars in Draco using data from 
the Sloan Digital Sky Survey (SDSS).  We present wide area, 
$u^*g^*r^*i^*z^*$ CCD observations of Draco down to about $g^* \cong 23$, and 
construct a color magnitude diagram.  We combine our catalog of SDSS stars 
in Draco with previously published photometric data in Draco, from: 
\cite{bs61}, \cite{cs86}, \cite{s97}, \cite{gmh98}, and \cite{p01} to 
create a heterogeneous catalog of Draco stars, with multiple epochs
of observation which allow us to identify variable candidates.

\section {OBSERVATIONS} 

The CCD imaging observations of Draco are taken from SDSS runs 1336 and 1339, 
which were obtained 2000 April 4 with the SDSS mosaic
imaging camera \citep{getal98}.  The majority of the stars in the
dwarf spheroidal fall on two frames of data in the same camera column
(one continuous stream of drift-scan data).  Since the precise
calibration for the SDSS filter system is still in progress,
magnitudes in this paper are quoted in the $u^* g^* r^* i^* z^* $
system, which approximates the final SDSS system \citep{setal2002}.  
These systems differ absolutely (with negligible color
terms) by only a few percent in $g^* r^* i^* z^*$, and no more than
10\% in $u^*$.  See \citet{figdss96} and \citet{yetal00} for further
information on the filter system and the overall survey, respectively.
The data were reduced with PHOTO (Lupton et al. in preparation)  version 5.1.7.  \citet{petal2002}
describes the astrometry and astrometric accuracy of the software.
\citet{hetal2002} describes our monitoring program for photometricity.  These data 
are included in the public Early Data Release (EDR) \citep{edr02}.

In the course of commissioning a survey, the images are reduced several
times as the software evolves.  The data presented here have been reduced
three times.  The first reduction is called ``rerun 0," and second is
``rerun 1."  Most of the data included in this paper are from ``rerun 1,"
with some exceptions which are noted in the text below.  ``Rerun 2"
was the version released in the EDR, and differs from ``rerun 1" only
in a small overall calibration offset.  Since we recalibrate all of
the photometry ourselves, the difference is not important.  So, an
object which is listed in the catalog by the SDSS id 1336-1-5-60-950
(Run-Rerun-CameraColumn-Field-ObjectID) is the same object
as 1336-2-5-60-950 in the EDR.

The seeing for the Draco scans was typically 1.8'' FWHM.  
Intercomparison of objects detected twice in overlapping scans is a
good indication of relative photometric error, and for objects with
$g^* < 19$, rms error for stellar sources is typically $<4$\%.  For
objects between $20 < g^* < 21$, typical errors are 8\%, growing to
20\% at $g^* = 23$, near the detection limit.  For reference, blue
stars with $0 < B-V < 0.2$ have a SDSS $g^*$ magnitude approximately
equal to their Johnson $V$ magnitude, while stars with $g^*-r^* = 1$
have $g\sim V+0.45$.  In general, $g^* = V + 0.54(B-V)-0.07$ \citep{setal2002}.

In order to obtain the best measurements at faint magnitudes, we used 
photometry derived from fits of the pointspread function (PSF) to stellar 
profiles.  The SDSS code determines the PSF as a function of position in 
each CCD frame (1361 x 2048 pixels, or $9^{'}$ x $13.7^{'}$).  While 
we were studying the stellar populations within Draco, we noticed a 
systematic shift in the color of the giant branch as a function of position 
within the Draco field.  Further analysis showed 
that this result could be explained by systematic inaccuracies in the 
(PSF) photometry as modeled in the EDR reductions of the data (these
systematics are corrected in later data releases).
Since the code used to create the EDR did not accurately track rapidly varying 
point-spread functions, the PSF magnitudes in some regions of the sky were
systematically shifted brighter or fainter by up to a tenth of a magnitude.

For this study, we were able to correct the EDR PSF magnitudes using the 
aperture magnitudes for bright objects.  Although aperture magnitudes 
are quite noisy for faint stars, 
they are accurate for bright stars and not affected by the error 
in the calculated PSF.  We used the brighter stars 
to calculate a correction to the PSF magnitudes, as a function of row and 
column in each frame.  We fit a second degree polynomial to the deviations 
between the aperture and PSF magnitudes (mag($_{AP}$)$-$mag($_{PSF}$)) vs. 
CCD row and column for stars brighter than magnitude 19.5.  We then used 
this polynomial to correct the photometry of {\it all} stars in the frame 
to the mean of the aperture photometry for that frame.  The assumption here 
is that the aperture magnitudes are correct, and their calibration does not 
vary as the PSF varies across rows or columns.  This is a good assumption 
if we capture practically all of the light in each object, even in the 
sections of data with poorer seeing.  

The results of this correction procedure are shown in 
Figure 1.  The left panel shows the color-magnitude diagram for Draco, using 
PSF magnitudes, and the right panel shows the same diagram after correcting 
the PSF magnitudes.  Though we did not use any
information about the giant branch of Draco, notice that the giant branch 
of the dwarf galaxy is significantly narrower after correction.  This is evidence
that the correction technique significantly improved the accuracy of the photometry.

\section {CREATING THE CATALOG}

In order to generate a useful catalog of stars in the Draco dwarf galaxy, we first selected
from the SDSS database those objects with $259^\circ < \alpha < 261^\circ$ and $57.4^\circ < \delta < 58.4^\circ$,
which were marked as unsaturated, and which were far enough from the edge of the frame
that they were completely contained on a single CCD detector.  This latter criterion 
effectively eliminates only quite extended galaxies near the edges, since there is
enough overlap between frames that point sources are completely contained on at least one frame.
We also removed duplicate measurements of the same astronomical object, either by multiple
observation or multiple software measurements of the same observation, by choosing only
observations marked in the SDSS as `primary.'  For each direction in the sky, only one observation and
instance of processing is considered `primary,' so this flag effectively removes duplicates.

To determine a centroid of our elliptical region, stars in our original data set 
were binned into boxes of area 0.05 degrees on a side.  Then, we fit an elliptical
Gaussian (right ascension and declination of the center, 
major and minor axes, a position angle, and sky), to the binned data.  
The adopted center of the distribution is:
$\alpha =  17:20:13.2,\>  \delta =  57:54:45$ (J2000).  The exact center
of the fit shifted by ten arcseconds depending on how we weighted the
fit, which yields an error bar on our chosen center.  Other sources such 
as \cite{bs61}, \cite{ih95}, and \cite{cca99} report the center of Draco as: 
$\alpha = $ 17:20:13 $\delta = $ 57:55:11 (J2000, converted from B1950), 
$\alpha = $ 17:20.3 $\delta = $ 57:55.1 (J2000, converted from B1950), and 
$\alpha = $ 17:20:12.39 $\delta = $ 57:54:55.3 (J2000) respectively.  
These positions for Draco's center are in good agreement with the value 
we found in right ascension, and off by about ten arcseconds in declination.  The 
major axis of the ellipse fit is nearly constant in declination (PA $\sim 90^\circ$)
with Gaussian sigma 0.122 degrees (corrected for $\cos(\delta)$).  The minor axis 
is 0.082 degrees in the declination direction implying an ellipticity of
1-0.082/0.122=0.33 in agreement with Hodge (1964) who found an ellipticity of $0.29\pm 0.04$, and also in excellent agreement with \citet{p01} who find an 
ellipticity of 0.331 for Draco.  

To increase the fraction of cataloged stars which are associated 
with the dwarf galaxy, we included only those stars in an 
elliptical region centered on the Draco dwarf galaxy (see Figure 2), which extends to about one
third of the tidal radius measured by \citet{oden01}.
Of the 7417 objects within the area of this 2-sigma
ellipse for our Draco catalog, 5492 were brighter
than our adopted magnitude limit of 23.005 (this odd effective limit is due to rounding
before the limit was applied).  The actual number of SDSS objects in the final catalog, 5634,
is slightly different, as 15 objects with suspicious SDSS photometry (mostly on bright
star bleed trails) were removed, and 157 objects were added from previous data
reductions of the same astronomical images.
Positions on the sky of cataloged SDSS stars are displayed in Figure 2.
Faint SDSS detections ($g^* > 22$) are shown as small black dots,
while brighter stars are larger black dots.  Galaxies (objects which the
SDSS software determined to be extended) are not included in the plot.

The catalog presented in this paper overlaps several previously published
photometric surveys of the Draco dwarf galaxy.  The footprints of the five
surveys which are cross-referenced to sources in our catalog are shown
in Figure 2.  Our catalog also includes the fainter ten of the twelve 
Draco stars with accurate photometry from \citet{hm2000}.  
Their SDSS IDs, from brightest to faintest, are:
1339-1-5-62-26, 1336-1-5-61-148, 1336-1-5-61-119, 1339-1-5-61-384, 1336-1-5-61-478,
1339-1-5-61-366, 1336-1-5-61-370, 1336-1-5-61-294, 1336-1-5-61-685, and 1336-1-5-61-1227.  
Since there are so few, we do not include a separate column in the catalog
table.  In the remainder of this section, we describe the
construction of a catalog of Draco stars with cross-references to the other
catalogs.  Ultimately, the overlapping catalogs are used to generate a list
of possible variable stars.

We list in Table 1 a summary of 
stars detected in each previous survey, along with numbers of matches to SDSS 
detections.  The main data product of this paper, with format specified in Table 2, 
is contained in the (electronic) Table 3.  Table 3 contains identifications, magnitudes, 
colors and cross-references to other surveys for stars in Draco.  The SDSS J2000 
right ascension (in degrees), declination (in degrees), SDSS id 
(Run-Rerun-CameraColumn-Field-ObjectID), $g^*$ magnitude, 
and $u^*-g^*, \> g^*-r^*, \> r^*-i^*,\> i^*-z^*$ colors of each object are listed 
as columns 1 through 8 of Table 3, respectively.  These magnitudes have been derreddened
using the E(B-V) tabulated in column 9, as derived from
\citet{sfd98}.  Following \citet{ccm89}, $A_{u^*}=5.155E(B-V)$, $A_{g^*}=3.793E(B-V)$, 
$A_{r^*}=2.751E(B-V)$, $A_{i^*}=2.086E(B-V),$ and $A_{z^*}=1.479E(B-V)$.  Measurements of 
the same objects in other
catalogs are listed in columns 10 through 33, as described below.  Note that the SDSS
photometry is dereddened, while that of all the other catalogs is not.
A flag indicating classification
of the object as stellar (``s"), a resolved galaxy (``g") or
of an indeterminant extent (``f" for faint) is indicated in column 34 of
Table 3.  All stars with $g^* > 22$ are classified as ``faint."
  The 5766 lines in this table include the 5634 individual objects detected in the
SDSS database, 19 objects from the Baade-Swope dataset which were not matched to SDSS
objects, and 113 duplicate entries in cases where there exist Piatek objects in
both the E1 and CO fields.  The additional Baade-Swope entries were apparent in the
SDSS images, so we were able to provide approximate astrometric positions even though
no automated photometry was generated.  We included multiple lines for Piatek duplicates
to avoid adding yet four more columns to an already large table.

\subsection {Photometry from \cite{bs61}}

The survey of area of \cite{bs61} (hereafter BS61), approximately covering 
a circular region of radius $5^{'}$$26^{''}$ centered at R.A. $17^h19^m24^s$, 
Dec. $+57^{\circ}$$58^{'}$$06^{''}$ (B1950),  
is completely contained within our catalog.  Since the individual 
right ascensions and declinations were not included for each star 
in the BS61 catalog, the stars were matched by comparing the finding 
charts with SDSS images.  
Of the 624 stars in Table C of BS61, 563 were cross-identified with an
object in the SDSS catalog, available in the SDSS Early Data Release (EDR).  An
additional 24 objects were measured in a previous software reduction (rerun 0) of the
same data, and were included in our catalog.  The majority of the data in Table
3 are from a software reduction called ``rerun 1."  The second number in the third field 
of Table 3 is the SDSS id rerun number, thus, a 0 there would indicate that 
rerun 0 was used.  The photometry for these latter
objects could differ systemmatically by a small amount from the majority of the
objects in the catalog, and were not corrected to match large aperture photometry of bright stars.  
Several other BS61 stars were apparent in the SDSS images, but were not reliably 
measured in any automated SDSS data reduction.  In most cases, the star was 
close enough to a very bright
star that the automated detection software had difficulty deblending it.  
The approximate positions for nineteen such objects are included in Table 3, though no
SDSS photometry is provided.  One star could not be identified 
because it was not labeled in the BS61 finding chart.  The remaining BS61 stars either could not
be reliably identified on the SDSS images or lay close enough to an image defect
that the data could not be used reliably.

The process of cross-referencing catalogs through visual identification 
of stars in finding charts and images is inherently prone to errors.  To 
reduce and evaluate the error rate, several checks were performed.
We plotted $g^*-V$ vs. $B-V$ for a set of magnitude ranges 
(Figures 3a and 3b).  Here, $g^*$ is the SDSS magnitude 
which is most similar to V.  Except in unusual cases, stellar photometry 
should produce an approximately linear relationship between $g^*-V$ and 
$B-V$.  The scatter should increase with the photometric errors at fainter 
magnitudes.  The derived color relations (SDSS dereddened values compared to BS61
undereddened values) are:
\begin{equation} g^*=V + 0.618(B-V) - 0.23, \end{equation}
\begin{equation} g^*-r^* = 1.05 (B-V) - 0.28. \end{equation}
This transformation is only slightly different from that found
in \cite{figdss96} (hereafter F96), based on preliminary SDSS filter
transmission design.
Outliers on the $g^*-V$ vs. $B-V$ plots were checked for mis-identification matches
between SDSS and BS61, and several were found and corrected.
Using the derived color transformations, the magnitudes of the detections
of the same objects were directly compared in magnitude
and color.  Stars with significant differences in computed magnitudes
or colors between different catalog observations are either variable, have
unusual spectra which do not lead to linear color transformations, or are
errors in either the SDSS or BS61 measurements.   The outliers
in magnitude are marked with diamonds in Figure 4a, and those outliers
in color by diamonds in Figure 5a.  Note that the photometry of BS61 is systematically
shifted by several tenths of a magnitude at the faint end.

We compared the photometry between catalogs only for objects which are classified in
the SDSS as stellar (column 34 of Table 3).  Since the fraction of the light measured
for galaxies differed between datasets, the inclusion of extended sources in the
outlier detection process led to a large number of objects which appeared to vary in brightness
from one observation to the next.  This data cut did not have a large effect on
the results from the BS61 data, since BS61 pre-selected only stellar objects, but is of 
greater import in comparing objects in other catalogs discussed below.  Of the 586 BS61 
objects matched to SDSS objects, only 25 were classified as galaxies by the SDSS, 
and 6 were fainter than $22^{nd}$ magnitude in $g^*$ (and thus are classified only as ``faint" in Table 3).

Stellar objects with a significant calculated photometric difference between 
SDSS and BS61 are flagged in column 13 of Table 3.  Stars without significant photometric 
differences (as seen in Figures 4a and 5a) are flagged 0, stars 
with significant differences, but which have magnitudes and colors 
consistent with RR Lyrae variables 
(selection criteria are discussed below) are flagged as 1 and other variable
candidate stars are flagged as 2 in this column.  A flag value of 3 denotes
an object which was not measured in the SDSS, BS61, or both.
The BS61 id, V magnitude, and $B-V$ values are found in columns 10, 
11, and 12 of Table 3, respectively.    

\subsection {Photometry from \cite{s79}}

The \cite{s79} (hereafter S79) photometric survey consisted of new 
photometry for a subset (512 stars) of the BS61 catalog.  The star identifications 
of the stars in Table V of S79 are the same star identifers as in Table C 
of BS61.  Since the S79 stars are identical to the BS61 stars, the S79 
photometric data were simply matched to the appropriate BS61 stars.  The S79 id, V magnitude, and $B-V$ values are found in columns 14, 15, and 16 
respectively in Table 3.  The photometric difference flag in column 17 of
Table 3 indicates whether or not there is a significant difference between the S79 (recalibrated
BS61) photometry and SDSS data, as graphically shown in Figures 3c and 3d.
The same $(B,B-V) \rightarrow (g,g-r)$ transform was performed as was used for
BS61, and the outliers are shown in Figures 4b and 5b.
The values of the flags have the same meanings as they do
for BS61 (0:no difference, 1:RR Lyrae candidate, 2:unknown variable
candidate, 3:no SDSS detection).

\subsection {Photometry from \cite{cs86}}

The survey area of \cite{cs86} (hereafter CS86) is broken up 
into two overlapping fields: Field 1 and Field 2.  The total area (27 
square arcminutes) of these two fields is completely contained within our 
catalog.  Each of the two fields covers $4^{'}$$.5$ in declination 
and $2^{'}$$.9$ in right ascension and an outline of the field 
boundaries is drawn in Figure 2.  Each field is approximately centered 
on a BS61 star; Field 1 is centered on BS69 star 92 and Field 2 on star 289.  
The two fields overlap by 3.4 square arcminutes (26\%).  An x and y value 
(in pixels) is given for each star in the two fields.  To convert 
these x and y values to right ascension and declination, we created a 
transformation for each field.  These transformations 
are accurate within about 1 arcsecond (rms).  The transformation for Field 1 is:

\begin{equation} \alpha = 260.08088 - 0.000307403*x - 0.00000397545*y \end{equation}
\begin{equation} \delta = 57.87482 - 0.00000211255*x + 0.000163354*y, \end{equation}

and for Field 2 the transformation is:

\begin{equation} \alpha = 260.0153958 - 0.00030713911*x - 0.00000316997*y \end{equation}
\begin{equation} \delta = 57.88494 - 0.0000016845*x + 0.0001632135*y. \end{equation}

About half of the stars in each of the two fields were faint
(V $\geq$ $23$), and were not matched to the SDSS data. 
370 of the 547 Field 1 stars were matched to SDSS detections, and 
292 of the 423 Field 2 stars were matched.
Most of the stars that were not identified had $V>22.5$).  Since the 
unmatched stars were not clumped in position
on the sky, we believe that the coordinate transformation is not
responsible for the fact that many of stars are not found in common
between the catalogs.  Almost all of the unmatched CS86 stars were either
fainter than 23rd magnitude in V or close enough to brighter stars that
the SDSS deblender did not identify them as isolated sources.

We used the same method to reduce and evaluate the error rate in CS86 data 
as we did in BS61 data.  The transformation from B,V photometry to $g^*,r^*$ photometry
was different, however.  The transformation from F96 was a good match to the data,
so it was adopted:
\begin{equation} g^* = V + 0.56 (B-V) -0.12, \end{equation}
\begin{equation} g^*-r^* = 1.05 (B-V) - 0.23. \end{equation}
The computed $g^*$ values would differ by a few percent if we used the transformations
derived in \citet{setal2002} from comparisons of SDSS standard 
stars to Landolt standards.  A comparison of our photometry with the corresponding
photometry from ten stars in \citet{hm2000} support the claim that the 
\citet{setal2002} transformations are a better match to modern CCD photometry
than the theoretical transformations of F96.  Since we are using the transformations
only to look for outliers, it doesn't matter if the overall transformation is 
off by a few percent (either due to the transformation or the fact that our
magnitudes are dereddened and CS86 data is not).

Figures 3e, 3f, 4c and 5c show the photometric comparisons between SDSS and CS86
data, and the outliers chosen as candidate variables.  Twenty
candidate variable objects were identified (of which 16 are consistent with
Draco RR Lyrae variables in color and magnitude).  These 
objects are either variable, have unusual spectra, or
highlight photometric errors in at least one catalog.
The CS86 Field 1 id, V magnitude, $B-V$ color, and variability flag are found in 
columns 18, 19, 20, and 21, respectively, and 
the CS86 Field 2 id, V magnitude, $B-V$ color, and variability flag are found in 
columns 22, 23, 24, and 25, respectively.

\subsection {Photometry from \cite{p01}}

Out of the nine tables in and around Draco in \cite{p01} (hereafter P01), 
only three (Tables a, b, and d) overlap data within our catalog.  
Field C0 from P01 contains the majority of stars found within our catalog.  
Of the 11381 C0 stars within our catalog region, 
4796 stars matched to stars within our catalog.
132 of the unidentified 
stars were found in the rerun 0 version of the reduction software.
As was true in comparing our catalog to those of BS61 and CS86,
most of the unmatched stars brighter than $V = 22.5$ were not detected in the SDSS
because they were not deblended from bright stars.
Out of the 1294 stars in E1 (P01 Table b) and 191 stars in W1 (P01 Table d) within 
our catalog region, 444 and 60, respectively, stars were identified in our catalog.
  An additional five of the unidentified stars in Table b and two of the
unidentified stars from Table d were found in the rerun 0 
version of the reduction software.  A total of 5326 matches to P01 objects were
made out of a total of 5634 unique objects in the SDSS catalog.  Again, about 
90\% of the stars brighter than $V = 22.5$ are matched with objects in the SDSS database.

For bluer colors, $V-R \le 0.875$, the F96 color transformations work well:
\begin{equation} g^* = V + 0.96 (V-R) - 0.14, \end{equation}
\begin{equation} g^*-r^* = 1.80 (V-R) - 0.27. \end{equation}
For redder stars, we used the transformation:
\begin{equation} g^* = V + 0.70, \end{equation}
\begin{equation} g^*-r^* = 0.03. \end{equation}
Figures 3g, 3h, 4d, and 5d show a comparison of the photometry between P01 and
SDSS catalogs; stars with inconsistent photometry are marked with a diamond-shaped symbol.
134 of the 2374 matched objects are flagged as possible 
variable stars.  Due to saturation of P01 photometry on the bright end, we
did not compare photometry for stellar objects brighter than $V=17$.  The P01 id, 
V magnitude, $V-R$ color, and match flag are found in columns 26--29 of Table 3, respectively.

Note that in cases where there is a match of the same SDSS star to
a P01 C0 and either a P01 E1 or P01 W1 field, that there are two
lines of entries in Table 3, one for each match, with the SDSS information
duplicated.

\subsection {Photometry from \cite{gmh98}}


The survey of area of the HST observations of Draco by \cite{gmh98} 
(hereafter G98) is completely contained within our catalog and is roughly 
centered and almost completely contained within CS86 Field 1 (See Figure 2).  It was 
necessary to 
create a transformation from G98 x and y to R.A. and Dec. for the three 
G98 fields (WF2, WF3, and WF4).  For the WF2 field the transformation is:
\begin{equation} \alpha = 260.038354 - 0.0000336621*x - 0.00003946221*y \end{equation}
\begin{equation} \delta = 57.90812 - 0.0000209702*x + 0.000017888*y. \end{equation}
For WF3 the transformation is:
\begin{equation} \alpha = 260.039295 - 0.000039238*x + 0.000033847*y \end{equation}
\begin{equation} \delta = 57.90852 + 0.0000179864*x + 0.000020851*y. \end{equation}
And for WF4 the transformation is:
\begin{equation} \alpha = 260.04099 + 0.0000342794*x + 0.000038953*y \end{equation}
\begin{equation} \delta = 57.90783 + 0.0000206995*x - 0.000018216*y. \end{equation}
These fields are from one pointing of WFPC2.

The G98 data are taken with the Hubble Space Telescope and therefore includes
very faint stars.  There are 50 stars with V magnitude $\leq$ 23 in the WF2 data, 
46 in the WF3 data, and 56 in the WF4 data.  A total of 88 stars 
were matched by position to stars in our catalog, including 6 from the ``rerun 0"
version of the catalog.  Since F96 provided no transformations from V,I to $g^*,r^*$
filter systems, we generated our own transformation equations from the data:
\begin{equation} g^* = V + 0.991 (V-I) - 0.15, \end{equation}
\begin{equation} g^*-r^* = 1.25 (V-I) - 0.24. \end{equation}
The G98 id (id-WF chip number), V magnitude, $V-I$ color, and photometric 
matching flag are found in columns 30, 31, 32 and 33 of Table 3, respectively.
No stars were selected as variables by comparison to the G98 catalog (see
Figures 3i, 3j, 4e and 5e).  

\section {ANALYSIS}

The catalog of Table 3 is intended to provide a reference to researchers
interested in the stellar populations and structure of Draco. The extensive
cross-references to the literature provide an opportunity to select variable
candidates based on multi-epoch photometry over a long baseline.  Though
the filter systems are inhomogeneous, roughly linear behavior for most
colors yields a good list of candidate variables across the Draco field.

Table 4 lists all 142 candidate variables selected from Table 3. The columns of 
Table 4 are: right ascension, declination, SDSS ID, SDSS $g^*$, $g^*-r^*$, $u^*-g^*$, 
suspected variable type (RR Lyrae , QSO, Cepheid or Unknown), matching catalog 
(BS61, S79, CS79, P01, or G98), and the $\delta g^*$ and $\delta (g^*-r^*)$ between 
the SDSS observation and the transformed magnitude and color of the cross
reference observation onto the SDSS $(g^*, g^*-r^*)$ system.  A positive value
of delta $g^*$ or delta $g^*-r^*$ means the star was fainter or redder respectively in the SDSS than in the comparison
catalog.  The last column gives the
identity of the object from other catalogs.  

Since the observations with the full set of filters for SDSS observations for
each object are obtained within minutes of each other, most stars can be
expected to be at a single point in their light curve.  By plotting the colors
of known RR Lyrae stars from BS61, we determined that in Draco these
variable stars lie in a narrow region of the H-R diagram (see Figure 6).  Therefore,
we tentatively identify all 112 stars in the range 
$19 < g^* < 20.8$, $g^*-r^* < 0.4$, $2.7(g^*-r^*)+19.2  
<  g^* < 2.7(g^*-r^*)+20.1$ as RR Lyrae candidates.  These objects have a one in
the variability flag fields of Table 3, and are labeled as ``RR Lyrae" in Table
4.  Objects which are suspected to vary but which do not lie in the region
inhabited by RR Lyrae stars have a one in the variability flag fields of Table 3
and are labeled as ``unknown" in Table 4.  There are only 30 objects suspected of
varying which have colors outside the region known to contain RR Lyrae variables.
A few of these might be RR Lyrae variables as well, as their colors place them only
slightly outside our RR Lyrae color box.

The location of all variable candidates is summarized in the color magnitude diagram 
presented in Figure 6.  
Variable candidates from each cross-referenced catalog are marked with 
a separate color while all SDSS stars are indicated as black dots.  
The variables from the BS61 survey are well confirmed \citep{n85} 
with full light curves.  We were able to find the positions of 135 of the 137 BS61
variable objects within eight arcminutes of the center of Draco.  Two 
of the 137 did not appear to be
in the central part in the finding charts, so they were not included.  Note that
the numbering system for the BS61 variable stars is separate from the photometric
table with which we matched our data; we did not match the photometry for the variable
star list.  We recover 82 (60\%) of the 135 BS61 variables, which are identified in the last
column of Table 4.  We also recover 28 numbered BS61 variables outside the inner eight arcminutes,
and three BS61 variables which were not given a number.  Since most of the variables were detected
from only two epochs of photometry, we would not expect to recover all of the variable
objects in Draco from this technique.

Two of the variable candidates (1339-1-5-60-207 and 1339-1-6-61-504) are known quasars from
\citet{schetal2002}.  These objects are indicated by a large black circle in Figure 6,
and labeled ``QSO" in the last column of Table 4.


Five known carbon stars are included in Table 3.  
SDSS stars 1336-1-5-61-267, 1336-1-5-61-436, and 
1336-1-5-61-484 correspond to \citet{als82} stars $C2=J$, C1, and C3, 
respectively.  SDSS stars 1336-1-5-60-249, 1336-1-5-61-436, and 1336-1-5-60-294
correspond to \citet{metal2002} stars J171942.4+575838, J171957.7+575005, and
J172038.8+575934, respectively.  Object 1336-1-5-61-267 is apparently variable and
we detect it as such in multiple catalog comparisons.  It appears as a multi-colored point 
at $(g^*, g^*-r^*) = (18.02, 1.13) $ in Figure 6, and is identified as type ``Carbon" in the
last column of Table 4.

\section {CONCLUSIONS}

A matched catalog of 5634 objects in the Draco dwarf field, 
extending over four times the area of the \citet{bs61} catalog and 
reaching to approximately one third the new tidal radius of \citet{oden01}, has 
been assembled.  We selected five color data from the SDSS with $15 < g^* < 23$,
within the elliptical region $[(\alpha-260.051)/0.460]^2 + [(\delta-57.913)/0.164]^2 < 1$, 
where $\alpha$ and $\delta$ are in J2000 degrees.
This SDSS data was merged with several existing data sets to produce our
final catalog. The data sets include: \citet{bs61}, \citet{cs86}, \citet{s97}, 
\citet{gmh98}, and \citet{p01}.  

Using only stellar sources with $g^*<23$, we identify
142 candidate variable stars, of which 113 were identified as variables in BS61, one
is a known carbon star, and two are known QSOs.  Since the SDSS observations
were taken nearly simultaneously in all filters, the catalog colors and
distance to the Draco dwarf can be used to identify potential RR Lyrae stars.
Nearly $80\%$ of the candidate variable objects have colors consistent with 
those of RR Lyrae variables.  Only 6 (23\%) of the 26 objects which do not have previous
identifications are classified as RR Lyrae candidates on the basis of their colors.  
We report all astrometric and photometric 
transformations used to compare our data with previous catalogs.

\section {ACKNOWLEDGMENTS}

We would like to thank Carl J. Grillmair for sending his unpublished data, and Hugh
Harris for providing us with the positions of BS61 variables outside of the central
eight arcminutes of Draco.  The referee, Slawomir Piatek, made many helpful comments which improved
the paper.

This research has made use of the SIMBAD database, operated at CDS, Strasbourg, France.

This work was supported by Fermi National Accelerator Laboratory,
under U.S. Government Contract No. DE-AC02-76CH03000.

Funding for the creation and distribution of the SDSS Archive has been provided by the 
Alfred P. Sloan Foundation, the Participating Institutions, the National Aeronautics 
and Space Administration, the National Science Foundation, the U.S. Department of Energy, 
the Japanese Monbukagakusho, and the Max Planck Society. 
The SDSS Web site is http://www.sdss.org/.

The SDSS is managed by the Astrophysical Research Consortium (ARC) for the Participating 
Institutions. The Participating Institutions are The University of Chicago, Fermilab, 
the Institute for Advanced Study, the Japan Participation Group, 
The Johns Hopkins University, Los Alamos National Laboratory, 
the Max-Planck-Institute for Astronomy (MPIA), the Max-Planck-Institute for Astrophysics (MPA), 
New Mexico State University, University of Pittsburgh, Princeton University, the 
United States Naval Observatory, and the University of Washington.

\clearpage

\figcaption {Draco photometry without (left) and with (right) photometric correction.  
The PSF photometry in the EDR contains systematic errors which shift the photometry
in any filter in some regions of the sky brighter or fainter by up to a tenth of a
magnitude.  By comparing the PSF photometry to large aperture photometry for bright
stars, we were able to calculate the photometric corrections to the PSF photometry
as a function of position on the sky.
Note the much tighter giant branch after the correction.  \label{fg1}}

\figcaption {Catalog Footprint.  We show the positions of all of the brighter stellar objects 
($g^* < 22$, larger black dots) and faint objects ($g^* > 22$, smaller black dots) in our 
catalog.  For convenience, we include large black circles in the positions of bright
stars from the Guide Star Catalog (Lasker et al 1988).  We do not show extended objects brighter 
than $g^* = 22$, since these show
clustering that is not correlated with the clustering of stars in the Draco dwarf spheroidal
galaxy.  The positions of candidate variable stars selected by comparisons with photometry
in previous catalogs are shown.  For reference, we also show the field of view of the five other
catalogs of stars in Draco with which we compare our results.  The field of view of the Stetson
(1997) catalog is identical to that of Baade and Swope (1961).  Field 1 of Carney and Seitzer
(1986) is to the left (higher RA) of their field 2.
 \label{fg0}}

%

\figcaption {Filter transformations and Outlier Identification.  Dashed
lines are the transformations found 
in Fukugita et. al. (1996), while the solid lines are our computed transformations.  
The Fukugita transformation fits the Carney \& Seitzer (1986) data, but we fitted our
own transformations to the previous catalogs even though they use similar filters.  The 
diamonds indicate those stars whose photometry was significantly different between the
catalogs.  \label{fg3}}

\figcaption {Magnitude differences between stars in each catalog.  We show the difference
between the $g^*$ magnitude from the SDSS catalog and the $g^*$ magnitude computed
from each of the five previous Draco dwarf catalogs.  Only stellar objects (brighter
than $g^* = 22$) which are in common between the SDSS and each of the other catalogs are
included.  Diamonds indicate a candidate variable star.  \label{fg3a}}

\figcaption {Color differences between stars in each catalog.  We show the difference
between the $g^*-r^*$ color from the SDSS catalog and the $g^*-r^*$ color computed
from each of the five previous Draco dwarf catalogs.  Only stellar objects (brighter
than $g^* = 22$) which are in common between the SDSS and each of the other catalogs are
included.  Diamonds indicate a candidate variable star.  \label{fg3b}}

\figcaption {Variable Candidate Color Magnitude Diagram.  This H-R diagram shows the
data from Table 4.  The objects which are variable candidates are indicated with a different
color for each comparison catalog.  Some objects are selected as variable in more than one
catalog comparison.  The color-magnitude selection box for RR Lyrae variable candidates is
indicated within the red lines.  The two known QSOs which were selected as variables are 
indicated by large black circles.  The known carbon star which was selected as variable
is the multi-colored point at $(g^*, g^*-r^*) = (18.02, 1.13)$. \label{fg4}}

\clearpage

\begin{deluxetable}{ccccccccc}
\tablecolumns{9}
\footnotesize
\tablecaption{Summary of Object Matching to Previous Catalogs}
\tablewidth{0pt}
\tablehead{
\colhead{} & \multicolumn{1}{c}{BS61} & \multicolumn{1}{c}{S79} & \multicolumn{2}{c}{CS86} & \multicolumn{3}{c}{P01} & \multicolumn{1}{c}{G98}\\ 
\colhead{} & 
\colhead{} & 
\colhead{} & 
\colhead{Field1} & \colhead{Field2} &
\colhead{Table a} & \colhead{Table b} & \colhead{Table d}}

\startdata

No. of stars\tablenotemark{*} & 624 & 512 & 547 & 423 & 11,381 & 1,294 & 191 & 152\\
Total Matched & 586 & 488 & 370 & 292 & 4,928 & 449 & 62 & 88\\
Re-Run 1 Stars& 562 & 467 & 363 & 288 & 4,796 & 444 & 60 & 82\\
Re-Run 0 Stars& 24 & 21 & 7 & 4 & 132 & 5 & 2 & 6\\
Possible Variables\tablenotemark{**} & 2 & 6 & 11 & 9 & 126 & 7 & 0 & 0\\

\enddata
\tablenotetext{*}{Number of stars contained within our survey area with $V\leq23$}
\tablenotetext{**}{Includes RR Lyrae and Unknown variables}

\end{deluxetable}

\clearpage

\begin{deluxetable}{lcl}
\tablecolumns{3}
\tabletypesize{\scriptsize}
\footnotesize
\tablecaption{Format of Catalog in Table 3}
\tablewidth{0pt}
\tablehead{\colhead{Column} & \colhead{Format} & \colhead{Description}}

\startdata

1 & 9.5f & Right Ascension (J2000 degrees)\\
2 & 8.5f & Declination (J2000 degrees)\\
3 & 17s & SDSS ID (Run-Rerun-Camcol-Field-ID)\\
4 & 7.2f & $g^*$ - dereddened, aperture corrected Luptitude from PSF-fitting\\
5 & 6.2f & $u^*-g^*$\\
6 & 6.2f & $g^*-r^*$\\
7 & 6.2f & $r^*-i^*$\\
8 & 6.2f & $i^*-z^*$\\
9 & 6.4f & E(B-V)\\
10 & 6s & Baade-Swope ID\\
11 & 6.2f & Baade-Swope V\\
12 & 6.2f & Baade-Swope B-V\\
13 & 1d & Baade-Swope flags \tablenotemark{*}\\
14 & 6s & Stetson ID\\
15 & 6.2f & Stetson V\\
16 & 6.2f & Stetson B-V\\
17 & 1d & Stetson flags \tablenotemark{*}\\
18 & 6s & Carney-Seitzer Field 1 ID\\
19 & 6.3f & Carney-Seitzer Field 1 V\\
20 & 6.3f & Carney-Seitzer Field 1 B-V\\
21 & 1d & Carney-Seitzer Field 1 flags \tablenotemark{*}\\
22 & 6s & Carney-Seitzer Field 2 ID\\
23 & 6.3f & Carney-Seitzer Field 2 V\\
24 & 6.3f & Carney-Seitzer Field 2 B-V\\
25 & 1d & Carney-Seitzer Field 2 flags \tablenotemark{*}\\
26 & 9s & Piatek ID\\
27 & 7.3f & Piatek V\\
28 & 5.2f & Piatek V-R\\
29 & 1d & Piatek flags \tablenotemark{*}\\
30 & 6s & Grillmair ID-Field\\
31 & 6.3f & Grillmair V\\
32 & 6.3f & Grillmair V-I\\
33 & 1d & Grillmair flags \tablenotemark{*}\\
34 & 1s & Type (s - star; g - galaxy; f - faint, $g^* > 22$)
\enddata
\tablenotetext{*}{0 - Match; 1 - RR Lyrae; 2 - Unknown variable; 3 - Unmatched}
\tablecomments{Single spaces are inserted between columns.}

\end{deluxetable}

\begin{deluxetable}{rrcrrrrrrrrrr}
\tablecolumns{13}
\tabletypesize{\scriptsize}
\footnotesize
\tablecaption{Cross-Referenced Draco Catalog}
\tablewidth{0pt}
\tablehead{
\colhead{RA} & \colhead{DEC} & \colhead{SDSS id} & \colhead{g*} & \colhead{u*-g*} & \colhead{g*-r*} & \colhead{r*-i*} & \colhead{i*-z*} & \colhead{E(B-V)} & \colhead{BS id} & \colhead{BS V} & \colhead{BS B-V} & \colhead{BS flag}\\
}
\startdata
259.59433 & 57.90571 &  1339-1-5-60-1510 &   22.98 &   1.42 &   1.51 &   0.86 &   0.43 & 0.0262 &  & & & 3 \\ 
259.59869 & 57.90565 &  1339-1-5-60-1511 &   22.94 &  -0.16 &  -0.06 &  -0.66 &   1.46 & 0.0262 &  & & & 3 \\
259.59895 & 57.92744 &   1339-1-5-60-291 &   19.08 &   1.38 &   2.66 &   1.06 &   0.61 & 0.0257 &  & & & 3 \\
259.59919 & 57.94193 &  1339-1-5-60-1252 &   22.51 &   0.89 &   0.51 &   0.61 &   0.29 & 0.0254 &  & & & 3 \\
259.59940 & 57.91351 &   1339-1-5-60-383 &   21.31 &   1.32 &   1.57 &   1.19 &   0.62 & 0.0261 &  & & & 3  
\enddata
\tablecomments{Table 3 is presented in its entirety in the electronic version of The Astrophysical Journal Supplements.  A portion is shown here for guidance regarding its form and content.  The full catalog contains 34 columns of information on 5766 stars.}
\end{deluxetable}

\begin{deluxetable}{cclcccccccc}
\tabletypesize{\scriptsize}

\tablecolumns{11}
\footnotesize
\tablecaption{Candidate Variable Objects}
\tablewidth{0pt}
\tablehead{
\colhead{} & \colhead{} & \colhead{} & \colhead{} & \colhead{} & \colhead{} & \colhead{} & \multicolumn{1}{c}{Other} & \colhead{} & \colhead{} & \multicolumn{1}{c}{Variable}\\
\colhead{R.A.} &
\colhead{Dec.} &
\colhead{SDSS ID} &
\colhead{g} & \colhead{g-r} &
\colhead{u-g} & \colhead{type} & \colhead{Catalog(s)\tablenotemark{\dag,c}} & \colhead{delta g} & \colhead{delta g-r} & \colhead{Refrence}\tablenotemark{\ddag}}

\startdata
259.64816 & 57.87486 & 1339-1-5-61-655 & 20.27 & -0.01 & 1.31 & Unknown  & P01 &   0.559 &  0.116 & ---\\
259.66009 & 57.87728 & 1339-1-5-61-652 & 19.91 & 0.15 & 1.09 & RR Lyrae & P01 &  -0.680 & -0.210 & ---\\
259.70073 & 57.92844 & 1339-1-5-60-102 & 20.69 & 0.43 & 2.33 & Unknown  & P01 &  -0.025 & -1.000 & ---\\
259.73549 & 57.98246 & 1339-1-5-60-239 & 20.73 & -0.88 & 0.91 & Unknown  & P01 &   0.849 &  0.326 & ---\\
259.74238 & 57.88235 & 1339-1-5-61-689 & 20.14 & 0.18 & 0.80 & RR Lyrae & P01 &   1.153 &  1.242 & BS-76\\
259.74469 & 57.88249 & 1339-1-5-61-688 & 19.91 & 0.08 & 0.95 & RR Lyrae & P01 &  -1.237 & -1.108 & BS-109\\
259.75711 & 58.00798 & 1339-1-5-60-207 & 18.70 & 0.08 & 0.20 & Unknown  & P01 &  -0.524 &  0.062 & QSO\\
259.77357 & 57.92739 & 1339-1-5-61-569 & 20.37 & 0.26 & 0.87 & RR Lyrae & P01 &   0.684 &  0.404 & BS-159\\
259.77654 & 57.83001 & 1339-1-5-61-369 & 19.21 & 0.14 & 0.95 & Unknown  & P01 &   1.486 &  1.220 & BS-134\\
259.78363 & 57.97636 & 1339-1-5-60-261 & 18.24 & -0.04 & 1.15 & Unknown  & P01 &  -0.414 & -0.184 & BS-157\\
259.79617 & 58.01210 & 1339-0-5-60-49 & 20.84 & 1.80 & 1.56 & Unknown  & P01 &   0.393 &  0.370 & ---\\
259.79988 & 57.91029 & 1339-1-5-61-626 & 20.41 & 0.19 & 1.17 & RR Lyrae & P01 &  -0.497 & -0.764 & BS-22\\
259.82323 & 57.89237 & 1339-1-5-61-704 & 19.61 & -0.03 & 1.15 & RR Lyrae & P01 &  -0.784 & -0.246 & BS-73\\
259.82668 & 57.87930 & 1339-1-5-61-750 & 20.32 & 0.14 & 1.10 & RR Lyrae & P01 &   0.702 &  0.716 & BS-110\\
259.84590 & 57.83288 & 1339-1-5-61-106 & 19.69 & 0.02 & 1.19 & RR Lyrae & P01 &  -1.281 & -0.952 & BS-137\\
259.84705 & 57.93216 & 1339-1-5-61-590 & 20.54 & 0.28 & 0.93 & RR Lyrae & P01 &   0.933 &  0.874 & BS-15\\
259.86091 & 57.89279 & 1339-1-5-61-732 & 20.45 & 0.31 & 1.05 & RR Lyrae & P01 &   0.414 &  0.472 & BS-107\\
259.87057 & 57.82140 & 1339-1-5-61-908 & 20.40 & 0.31 & 0.90 & RR Lyrae & P01 &   0.410 &  0.364 & BS-47\\
259.87712 & 57.94266 & 1339-1-5-61-183 & 19.38 & -0.01 & 1.14 & RR Lyrae & P01 &   0.517 &  0.980 & BS-14\\
259.89192 & 57.92656 & 1339-1-5-61-235 & 20.43 & 0.17 & 1.18 & RR Lyrae & P01 &   0.672 &  0.602 & BS-18\\
259.89579 & 57.84657 & 1339-1-5-61-884 & 19.98 & 0.12 & 1.20 & RR Lyrae & P01 &   1.027 &  1.254 & BS-45\\
259.90025 & 57.90432 & 1339-1-5-61-280 & 18.12 & 0.19 & 1.26 & Unknown  & P01 &   0.328 &  0.388 & BS-194\\
259.90181 & 57.82497 & 1339-1-5-61-911 & 20.38 & 0.09 & 1.02 & Unknown  & P01 &   0.405 &  0.252 & BS-46\\
259.90992 & 57.79015 & 1339-1-5-62-406 & 20.28 & 0.16 & 1.11 & RR Lyrae & P01 &   0.413 &  0.520 & BS-50\\
259.91930 & 57.97742 & 1336-1-5-60-737 & 20.29 & 0.43 & 1.29 & Unknown  & P01 &   1.000 &  1.960 & ---\\
          &          &                  &       &       &       &          & S79\tablenotemark{*}   & -0.035 & -0.043 & \\
          &          &                  &       &       &       &          & BS61\tablenotemark{*}   &  0.008 & -0.169 & \\
259.92261 & 57.89095 & 1339-1-5-61-763 & 20.38 & 0.22 & 1.37 & RR Lyrae & P01 &   0.397 &  0.490 & BS-129\\
          &          &                  &       &       &       &          & CS86\tablenotemark{2}   &  0.455 &  0.811 & \\
259.92724 & 58.05734 & 1336-1-5-60-578 & 19.53 & -0.17 & 1.31 & RR Lyrae & P01 &  -1.572 & -1.600 & BS-2\\
259.92756 & 57.89161 & 1339-1-5-61-762 & 19.89 & 0.09 & 0.94 & RR Lyrae & P01 &  -0.541 & -0.234 & BS-188\\
          &          &                  &       &       &       &          & CS86\tablenotemark{2}   & -0.585 & -0.178 & \\
259.93739 & 57.90498 & 1336-1-5-60-525 & 20.28 & 0.28 & 1.04 & RR Lyrae & P01 &   0.399 &  0.298 & BS-127\\
          &          &                  &       &       &       &          & CS86\tablenotemark{2}   &  0.363 &  0.508 & \\
          &          &                  &       &       &       &          & S79   &  0.278 & -0.393 & \\
          &          &                  &       &       &       &          & BS61\tablenotemark{*}   &  0.318 & -0.487 & \\
259.93800 & 57.97613 & 1336-1-5-60-769 & 20.24 & 0.27 & 1.22 & RR Lyrae & P01 &   1.858 & -1.160 & ---\\
          &          &                  &       &       &       &          & S79\tablenotemark{*}   & -0.006 & -0.035 & \\
          &          &                  &       &       &       &          & BS61\tablenotemark{*}   &  0.038 & -0.067 & \\
259.94096 & 57.94053 & 1336-1-5-60-355 & 20.10 & 0.19 & 1.40 & RR Lyrae & P01 &   0.381 &  0.730 & BS-104\\
          &          &                  &       &       &       &          & CS86\tablenotemark{2}   &  0.243 &  0.614 & \\
259.94262 & 57.79564 & 1339-1-5-62-407 & 20.13 & 0.23 & 1.06 & RR Lyrae & P01 &   0.413 &  0.482 & BS-86\\
259.95101 & 57.91426 & 1336-1-5-60-1053 & 19.79 & 0.00 & 1.26 & RR Lyrae & P01 &  -0.308 &  0.090 & BS-20\\
          &          &                  &       &       &       &          & CS86\tablenotemark{2}   & -0.505 & -1.158 & \\
259.95303 & 57.94910 & 1336-1-5-60-867 & 20.42 & 0.23 & 1.24 & RR Lyrae & P01 &   1.075 &  0.914 & BS-147\\
          &          &                  &       &       &       &          & CS86\tablenotemark{2}   &  0.373 & -0.044 & \\
259.95442 & 57.99014 & 1336-1-5-60-1504 & 21.79 & 0.61 & 0.61 & Unknown  & P01 &   0.615 &  0.934 & ---\\
259.95470 & 58.03670 & 1336-1-5-60-622 & 20.59 & 0.37 & 1.16 & RR Lyrae & P01 &   0.514 &  0.586 & BS-93\\
259.95796 & 57.81791 & 1339-1-5-61-958 & 19.77 & -0.02 & 1.17 & RR Lyrae & P01 &  -0.926 & -0.812 & BS-48\\
259.96361 & 57.83811 & 1336-1-5-61-101 & 21.18 & 0.98 & 0.37 & Unknown  & P01 &   1.081 &  1.178 & ---\\
259.96370 & 57.81209 & 1339-1-5-61-967 & 20.09 & 0.11 & 1.02 & RR Lyrae & P01 &   0.671 &  0.632 & BS-85\\
259.96382 & 57.88907 & 1336-1-5-61-277 & 20.46 & 0.22 & 1.54 & RR Lyrae & P01 &   0.482 &  0.508 & BS-196\\
          &          &                  &       &       &       &          & CS86\tablenotemark{2}   &  0.726 &  0.905 & \\
259.96792 & 57.98671 & 1336-1-5-60-245 & 21.79 & 0.66 & 0.92 & Unknown  & P01 &   0.905 &  1.452 & BS-144\\
259.96842 & 57.87396 & 1336-1-5-61-993 & 20.42 & 0.22 & 1.18 & RR Lyrae & P01 &  -0.284 & -0.446 & BS-33\\
259.97258 & 57.81262 & 1339-1-5-61-969 & 20.37 & 0.23 & 0.82 & RR Lyrae & P01 &  -0.752 & -1.200 & BS-84\\
259.97731 & 57.88219 & 1336-1-5-61-946 & 20.42 & 0.23 & 0.96 & RR Lyrae & P01 &   0.886 &  0.752 & BS-78\\
259.98065 & 57.86038 & 1336-1-5-61-1078 & 20.39 & 0.20 & 1.23 & RR Lyrae & P01 &  -0.802 & -1.230 & BS-39\\
259.98082 & 57.81681 & 1336-1-5-61-479 & 19.55 & 0.04 & 1.16 & RR Lyrae & P01 &  -0.311 &  0.130 & BS-114\\
259.98440 & 57.87989 & 1336-1-5-61-328 & 20.08 & 0.17 & 1.21 & RR Lyrae & P01 &   0.609 &  0.926 & BS-32\\
259.98606 & 57.93344 & 1336-1-5-60-973 & 20.51 & 0.30 & 1.16 & RR Lyrae & P01\tablenotemark{*} &   0.129 &  0.120 & BS-16\\
          &          &                  &       &       &       &          & CS86\tablenotemark{1,*}   &  0.062 & -0.077 & \\
          &          &                  &       &       &       &          & CS86\tablenotemark{2}   &  0.573 &  0.763 & \\
259.99224 & 57.91310 & 1336-1-5-61-6 & 20.36 & 0.41 & 1.45 & Unknown  & P01\tablenotemark{*} &   0.014 & -0.148 & ---\\
          &          &                  &       &       &       &          & CS86\tablenotemark{1}   &  0.331 &  0.045 & \\
          &          &                  &       &       &       &          & CS86\tablenotemark{2,*}   &  0.176 &  0.003 & \\
259.99545 & 57.90403 & 1336-1-5-61-826 & 19.99 & 0.18 & 1.18 & RR Lyrae & G98\tablenotemark{*} &   0.125 &  0.150 & BS-24\\
          &          &                  &       &       &       &          & P01   &  0.216 &  0.486 & \\
          &          &                  &       &       &       &          & CS86\tablenotemark{1}   & -0.507 & -0.095 & \\
          &          &                  &       &       &       &          & CS86\tablenotemark{2,*}   & -0.265 &  0.286 & \\
259.99668 & 57.94619 & 1336-1-5-60-922 & 20.47 & 0.28 & 1.11 & RR Lyrae & P01 &   1.008 &  0.874 & BS-65\\
          &          &                  &       &       &       &          & CS86\tablenotemark{1}   &  0.439 &  1.012 & \\
          &          &                  &       &       &       &          & CS86\tablenotemark{2,*}   & -0.004 & -0.118 & \\
259.99791 & 58.05947 & 1336-1-5-60-597 & 20.38 & 0.24 & 0.97 & RR Lyrae & P01 &   1.210 &  1.032 & BS-55\\
259.99839 & 57.91189 & 1336-1-5-61-7 & 20.11 & 0.33 & 1.15 & RR Lyrae & G98\tablenotemark{*} &  -0.124 & -0.051 & ---\\
          &          &                  &       &       &       &          & P01   & -0.338 & -0.426 & \\
          &          &                  &       &       &       &          & CS86\tablenotemark{1,*}   & -0.081 &  0.046 & \\
          &          &                  &       &       &       &          & CS86\tablenotemark{2,*}   & -0.058 &  0.072 & \\
          &          &                  &       &       &       &          & S79\tablenotemark{*}   & -0.046 & -0.070 & \\
          &          &                  &       &       &       &          & BS61\tablenotemark{*}   &  0.009 & -0.154 & \\
260.00124 & 57.89062 & 1336-1-5-61-915 & 20.45 & 0.23 & 1.14 & RR Lyrae & P01 &   1.623 &  1.472 & BS-74\\
          &          &                  &       &       &       &          & CS86\tablenotemark{1,*}   & -0.061 & -0.116 & \\
          &          &                  &       &       &       &          & CS86\tablenotemark{2}   &  0.299 &  0.524 & \\
260.00285 & 57.89625 & 1336-1-5-61-267 & 18.02 & 1.13 & 3.07 & Unknown  & P01 &   0.319 &  0.212 & Carbon\\
          &          &                  &       &       &       &          & CS86\tablenotemark{1,*}   & -0.049 & -0.256 & \\
          &          &                  &       &       &       &          & CS86\tablenotemark{2,*}   & -0.025 & -0.248 & \\
          &          &                  &       &       &       &          & S79   & -0.128 & -0.362 & \\
          &          &                  &       &       &       &          & BS61\tablenotemark{*}   & -0.073 & -0.352 & \\
260.00939 & 58.02230 & 1336-1-5-60-476 & 19.58 & -0.01 & 1.17 & RR Lyrae & P01 &  -1.622 & -1.440 & BS-59\\
260.01286 & 57.83103 & 1336-1-5-61-1214 & 20.31 & 0.17 & 1.33 & RR Lyrae & P01 &  -0.336 & -0.442 & BS-113\\
260.02392 & 57.93595 & 1336-1-5-60-373 & 20.94 & 0.96 & 0.35 & Unknown  & P01 &   0.588 &  0.636 & ---\\
          &          &                  &       &       &       &          & CS86\tablenotemark{1}   &  0.615 &  0.695 & \\
          &          &                  &       &       &       &          & S79   &  0.660 &  0.665 & \\
          &          &                  &       &       &       &          & BS61   &  0.729 &  0.476 & \\
260.02500 & 57.89691 & 1336-1-5-61-884 & 19.61 & 0.08 & 1.24 & RR Lyrae & P01 &   0.054 &  0.512 & BS-27\\
          &          &                  &       &       &       &          & CS86\tablenotemark{1}   & -0.652 & -0.176 & \\
260.02551 & 58.03505 & 1336-1-5-60-641 & 19.96 & 0.11 & 1.11 & RR Lyrae & P01 &   0.345 &  0.560 & BS-58\\
260.02727 & 57.82912 & 1336-1-5-61-1242 & 20.61 & 0.30 & 1.15 & RR Lyrae & P01 &   0.376 &  0.390 & BS-186\\
260.02962 & 57.99697 & 1336-1-5-60-734 & 20.29 & 0.23 & 1.02 & RR Lyrae & P01 &   0.449 &  0.482 & BS-98\\
260.03149 & 57.97373 & 1336-1-5-60-290 & 18.47 & 0.70 & 1.58 & Unknown  & P01 &   0.374 &  0.844 & ---\\
          &          &                  &       &       &       &          & S79\tablenotemark{*}   & -0.051 & -0.047 & \\
          &          &                  &       &       &       &          & BS61\tablenotemark{*}   &  0.003 &  0.111 & \\
260.03395 & 58.04203 & 1336-1-5-60-637 & 20.30 & 0.37 & 1.19 & RR Lyrae & P01 &   0.631 &  0.712 & BS-57\\
260.03553 & 57.79121 & 1336-1-5-61-1389 & 20.29 & 0.27 & 1.09 & RR Lyrae & P01 &   0.517 &  0.612 & BS-49\\
260.03697 & 57.93967 & 1336-1-5-60-375 & 20.44 & 0.22 & 1.25 & RR Lyrae & P01 &   1.380 &  1.192 & BS-126\\
          &          &                  &       &       &       &          & CS86\tablenotemark{1,*}   & -0.036 &  0.104 & \\
260.03703 & 57.92476 & 1336-1-5-60-1097 & 20.48 & 0.31 & 1.05 & RR Lyrae & G98\tablenotemark{*} &  -0.097 & -0.053 & BS-160\\
          &          &                  &       &       &       &          & P01\tablenotemark{*}   & -0.099 &  0.004 & \\
          &          &                  &       &       &       &          & CS86\tablenotemark{1}   &  0.467 &  0.262 & \\
260.03847 & 57.91065 & 1336-1-5-61-250 & 20.05 & 0.20 & 1.11 & RR Lyrae & G98\tablenotemark{*} &  -0.121 &  0.025 & BS-68\\
          &          &                  &       &       &       &          & P01   & -0.839 & -1.024 & \\
          &          &                  &       &       &       &          & CS86\tablenotemark{1,*}   &  0.003 &  0.034 & \\
260.04789 & 57.96733 & 1336-1-5-60-860 & 20.42 & 0.18 & 1.02 & RR Lyrae & P01 &   0.970 &  0.972 & BS-123\\
260.05178 & 57.90319 & 1336-1-5-61-868 & 19.59 & -0.06 & 1.26 & RR Lyrae & P01 &  -0.773 & -0.258 & BS-72\\
          &          &                  &       &       &       &          & CS86\tablenotemark{1,*}   & -0.129 &  0.258 & \\
260.05709 & 57.95686 & 1336-1-5-60-915 & 20.36 & 0.23 & 1.36 & RR Lyrae & P01 &  -0.766 & -0.814 & BS-184\\
          &          &                  &       &       &       &          & CS86\tablenotemark{1,*}   &  0.187 &  0.011 & \\
260.06165 & 57.90899 & 1336-1-5-61-627 & 20.24 & 0.67 & 1.15 & Unknown  & G98\tablenotemark{*} &   0.218 &  0.130 & ---\\
          &          &                  &       &       &       &          & P01\tablenotemark{*}   &  0.225 &  0.220 & \\
          &          &                  &       &       &       &          & CS86\tablenotemark{1}   &  0.210 &  0.219 & \\
          &          &                  &       &       &       &          & S79\tablenotemark{*}   &  0.166 &  0.123 & \\
          &          &                  &       &       &       &          & BS61\tablenotemark{*}   &  0.320 &  0.018 & \\
260.06334 & 57.98805 & 1336-1-5-60-807 & 20.57 & 0.24 & 1.07 & RR Lyrae & P01 &  -0.373 & -0.570 & BS-8\\
260.06422 & 57.89098 & 1336-1-5-61-334 & 20.69 & 0.24 & 1.61 & RR Lyrae & P01\tablenotemark{*} &   0.124 & -0.300 & BS-171\\
          &          &                  &       &       &       &          & CS86\tablenotemark{1}   &  0.503 & -0.035 & \\
260.06524 & 57.90470 & 1336-1-5-61-875 & 20.54 & 0.32 & 0.93 & RR Lyrae & G98\tablenotemark{*} &  -0.010 &  0.075 & BS-71\\
          &          &                  &       &       &       &          & P01   &  0.757 &  0.644 & \\
          &          &                  &       &       &       &          & CS86\tablenotemark{1}   &  0.396 &  0.736 & \\
260.07075 & 57.87772 & 1336-1-5-61-1064 & 20.51 & 0.24 & 1.15 & RR Lyrae & P01\tablenotemark{*} &  -0.227 & -0.246 & BS-34\\
          &          &                  &       &       &       &          & CS86\tablenotemark{1}   &  0.645 &  0.030 & \\
260.07115 & 57.77800 & 1336-1-5-61-1423 & 19.95 & 0.15 & 1.35 & RR Lyrae & P01 &  -0.351 &  0.006 & BS-140\\
260.07345 & 57.86679 & 1336-1-5-61-91 & 20.15 & 0.20 & 1.23 & RR Lyrae & P01 &  -0.801 & -0.952 & BS-36\\
260.07361 & 57.95369 & 1336-1-5-60-950 & 21.87 & 0.51 & 0.93 & Unknown  & P01\tablenotemark{*} &   0.114 &  0.114 & ---\\
          &          &                  &       &       &       &          & CS86\tablenotemark{1,*}   & -0.020 &  0.173 & \\
          &          &                  &       &       &       &          & S79\tablenotemark{*}   &  0.117 & -0.016 & \\
          &          &                  &       &       &       &          & BS61   &  0.470 &  0.047 & \\
260.07414 & 57.95210 & 1336-1-5-60-948 & 19.22 & 0.21 & 1.04 & Unknown  & P01 &  -0.376 & -0.384 & BS-141\\
          &          &                  &       &       &       &          & CS86\tablenotemark{1}   & -0.308 & -0.078 & \\
260.07524 & 57.87247 & 1336-1-5-61-92 & 21.25 & 0.10 & 1.37 & Unknown  & P01 &   0.324 &  0.640 & ---\\
260.08537 & 58.01548 & 1336-1-5-60-714 & 20.64 & 0.27 & 1.02 & RR Lyrae & P01 &  -0.207 & -0.432 & BS-5\\
260.08782 & 57.87185 & 1336-1-5-61-1106 & 20.21 & 0.24 & 1.17 & RR Lyrae & P01 &   0.386 &  0.258 & BS-81\\
          &          &                  &       &       &       &          & S79   & -0.356 & -0.422 & \\
          &          &                  &       &       &       &          & BS61\tablenotemark{*}   & -0.289 & -0.653 & \\
260.09410 & 57.78254 & 1336-1-5-61-1420 & 21.28 & 0.65 & 0.51 & Unknown  & P01 &  -0.322 & -0.574 & ---\\
260.10083 & 57.86138 & 1336-1-5-61-1150 & 19.88 & 0.11 & 1.14 & RR Lyrae & P01 &  -0.579 & -0.124 & BS-132\\
260.10812 & 57.94047 & 1336-0-5-60-421 & 20.07 & 0.43 & 0.96 & Unknown  & P01\tablenotemark{*} &   0.059 &  0.142 & ---\\
          &          &                  &       &       &       &          & S79   & -0.332 & -0.169 & \\
          &          &                  &       &       &       &          & BS61\tablenotemark{*}   & -0.188 & -0.348 & \\
260.11124 & 57.90755 & 1336-1-5-61-290 & 20.51 & 0.20 & 1.20 & RR Lyrae & P01 &  -0.550 & -0.934 & BS-106\\
260.11766 & 57.95019 & 1336-1-5-60-417 & 19.85 & 0.06 & 1.04 & RR Lyrae & P01 &   0.287 &  0.672 & BS-103\\
260.12558 & 57.84441 & 1336-1-5-61-139 & 19.73 & 0.01 & 1.08 & RR Lyrae & P01 &   0.101 &  0.334 & BS-135\\
260.12974 & 57.96027 & 1336-1-5-60-958 & 20.43 & 0.29 & 1.06 & RR Lyrae & P01 &   0.435 &  0.308 & BS-158\\
260.13549 & 57.91925 & 1336-1-5-61-32 & 20.63 & 0.36 & 0.98 & RR Lyrae & P01 &   1.466 &  1.368 & BS-21\\
          &          &                  &       &       &       &          & S79   & -0.058 &  0.538 & \\
          &          &                  &       &       &       &          & BS61\tablenotemark{*}   & -0.025 &  0.506 & \\
260.13768 & 57.95849 & 1336-1-5-60-986 & 19.51 & -0.12 & 1.29 & RR Lyrae & P01 &  -0.884 & -0.156 & BS-124\\
260.14471 & 57.89721 & 1336-1-5-61-1016 & 21.94 & 1.54 & 1.38 & Unknown  & P01\tablenotemark{*} &   0.121 &  0.110 & ---\\
          &          &                  &       &       &       &          & BS61   &  0.498 &  0.342 & \\
260.14936 & 57.82179 & 1336-1-5-61-1346 & 20.44 & 0.45 & 0.83 & Unknown  & P01 &   0.735 &  0.450 & BS-138\\
260.15351 & 57.87019 & 1336-1-5-61-660 & 20.23 & 0.19 & 1.06 & RR Lyrae & P01 &   0.529 &  0.478 & BS-40\\
260.15828 & 57.92527 & 1336-1-5-61-848 & 20.32 & 0.32 & 1.46 & RR Lyrae & P01 &   1.040 &  0.932 & BS-162\\
260.15994 & 57.87658 & 1336-1-5-61-408 & 20.44 & 0.29 & 1.27 & RR Lyrae & P01 &  -0.364 & -0.430 & BS-35\\
260.16906 & 57.91449 & 1336-1-5-61-936 & 20.50 & 0.26 & 0.93 & RR Lyrae & P01 &   0.587 &  0.530 & BS-161\\
260.17357 & 57.83108 & 1336-1-5-61-1332 & 20.01 & 0.09 & 1.14 & RR Lyrae & P01 &   0.819 &  1.026 & BS-82\\
260.17433 & 57.96383 & 1336-1-5-60-996 & 20.48 & 0.25 & 0.83 & RR Lyrae & P01 &  -0.439 & -0.524 & BS-12\\
260.17624 & 57.98114 & 1336-1-5-60-895 & 20.38 & 0.21 & 0.93 & RR Lyrae & P01 &   0.745 &  0.768 & BS-169\\
260.17851 & 57.85796 & 1336-1-5-61-1235 & 19.64 & 0.02 & 1.11 & RR Lyrae & P01 &   0.224 &  0.758 & BS-41\\
260.18800 & 57.85765 & 1336-1-5-61-1239 & 19.58 & 0.02 & 1.04 & RR Lyrae & P01 &  -0.434 & -0.088 & BS-164\\
260.19653 & 57.96645 & 1336-1-5-60-999 & 19.31 & -0.17 & 1.16 & RR Lyrae & P01 &  -0.100 &  0.550 & BS-13\\
260.19697 & 57.92301 & 1336-1-5-61-902 & 19.92 & 0.04 & 1.29 & RR Lyrae & P01 &  -0.533 & -0.140 & BS-185\\
260.20741 & 57.90137 & 1336-1-5-61-1058 & 20.42 & 0.21 & 1.06 & RR Lyrae & P01 &  -0.547 & -0.780 & BS-25\\
260.21321 & 57.86324 & 1336-1-5-61-130 & 20.53 & 0.38 & 0.85 & RR Lyrae & P01 &   1.061 &  0.848 & BS-133\\
260.21460 & 57.97171 & 1336-1-5-60-509 & 19.26 & 0.63 & 1.46 & Unknown  & P01 &   0.821 &  1.548 & ---\\
260.21700 & 58.02371 & 1336-1-5-60-767 & 19.57 & -0.01 & 1.22 & RR Lyrae & P01 &  -1.409 & -0.946 & BS-94\\
260.21794 & 57.92034 & 1336-1-5-61-964 & 20.22 & 0.24 & 0.31 & RR Lyrae & P01 &  -0.377 &  0.258 & ---\\
          &          &                  &       &       &       &          & BS61\tablenotemark{*}   &  0.253 &  0.344 & \\
260.21987 & 57.84490 & 1336-1-5-61-159 & 19.61 & -0.01 & 1.42 & RR Lyrae & P01 &   0.441 &  0.944 & BS-177\\
260.22526 & 57.95146 & 1336-1-5-61-777 & 20.71 & 0.83 & 0.79 & Unknown  & P01 &   0.342 &  0.434 & ---\\
260.23602 & 57.89786 & 1336-1-5-61-1100 & 20.53 & 0.28 & 1.33 & RR Lyrae & P01 &   1.297 &  1.144 & BS-75\\
260.23839 & 57.83363 & 1336-1-5-61-1361 & 19.48 & -0.05 & 1.10 & RR Lyrae & P01 &  -1.615 & -1.480 & BS-178\\
260.24357 & 57.89214 & 1336-1-5-61-649 & 20.42 & 0.29 & 0.97 & RR Lyrae & P01 &   1.051 &  1.082 & BS-175\\
260.24627 & 58.00153 & 1336-1-5-60-858 & 19.78 & -0.07 & 1.28 & RR Lyrae & P01 &  -0.708 & -0.574 & BS-97\\
260.24700 & 58.02391 & 1336-1-5-60-270 & 19.47 & -0.03 & 1.16 & RR Lyrae & P01 &  -0.079 &  0.420 & BS-118\\
260.25222 & 57.86466 & 1336-1-5-61-1267 & 21.37 & -0.13 & 0.61 & Unknown  & P01 &   0.579 &  0.392 & ---\\
260.26119 & 57.88066 & 1336-1-5-61-673 & 19.76 & 0.07 & 1.10 & RR Lyrae & P01 &   0.734 &  0.988 & BS-80\\
260.26461 & 57.99732 & 1336-1-5-60-888 & 20.36 & 0.23 & 1.04 & RR Lyrae & P01 &   0.796 &  0.896 & BS-96\\
260.27663 & 57.86468 & 1336-1-5-61-1285 & 20.06 & 0.11 & 1.20 & RR Lyrae & P01 &   1.021 &  1.154 & BS-112\\
260.27901 & 57.99500 & 1336-1-5-60-913 & 20.46 & 0.28 & 1.07 & RR Lyrae & P01 &  -0.325 & -0.620 & BS-95\\
260.27955 & 57.90240 & 1336-1-5-61-396 & 20.39 & 0.24 & 1.06 & RR Lyrae & P01 &  -0.327 & -0.696 & BS-70\\
260.28048 & 57.96674 & 1336-1-5-60-1073 & 20.34 & 0.16 & 1.23 & RR Lyrae & P01 &   1.171 &  1.042 & BS-183\\
260.28602 & 57.79628 & 1336-1-5-62-310 & 20.14 & 0.18 & 1.24 & RR Lyrae & P01 &   0.295 &  0.540 & BS-87\\
260.30299 & 58.02519 & 1339-1-6-61-405 & 19.94 & 0.03 & 1.34 & RR Lyrae & P01 &   0.416 &  0.750 & BS-181\\
260.30423 & 57.89739 & 1336-1-5-61-1147 & 19.65 & 0.02 & 1.04 & RR Lyrae & P01 &  -0.437 &  0.146 & BS-189\\
260.30846 & 57.90981 & 1336-1-5-61-1102 & 20.33 & 0.25 & 0.94 & RR Lyrae & P01 &  -0.221 & -0.470 & BS-128\\
260.31889 & 57.89216 & 1336-1-5-61-1178 & 19.98 & 0.14 & 1.20 & RR Lyrae & P01 &  -0.959 & -0.904 & BS-26\\
260.32756 & 57.79548 & 1336-1-5-62-319 & 19.84 & 0.10 & 1.21 & RR Lyrae & P01 &  -0.497 & -0.026 & BS-199\\
260.32928 & 57.89011 & 1336-1-5-61-1193 & 20.35 & 0.20 & 0.91 & RR Lyrae & P01 &   0.897 &  0.902 & BS-130\\
260.33418 & 57.82171 & 1336-1-5-61-1426 & 20.35 & 0.13 & 1.07 & RR Lyrae & P01 &   0.731 &  0.670 & BS-166\\
260.34115 & 57.82410 & 1336-1-5-61-1422 & 20.16 & 0.05 & 1.17 & RR Lyrae & P01 &   0.456 &  0.194 & BS-139\\
260.35674 & 57.84176 & 1336-1-5-61-190 & 20.44 & 0.35 & 1.09 & RR Lyrae & P01 &   0.764 &  0.818 & BS-197\\
260.37540 & 57.97799 & 1339-1-6-61-469 & 20.24 & 0.22 & 1.06 & RR Lyrae & P01 &   0.450 &  0.400 & BS-61\\
260.39565 & 57.85005 & 1339-1-6-62-330 & 20.79 & 0.78 & 0.98 & Unknown  & P01 &   0.508 &  0.150 & ---\\
260.40182 & 57.93293 & 1339-1-6-61-551 & 20.19 & 0.14 & 1.18 & RR Lyrae & P01 &   0.482 &  0.338 & ---\\
260.41938 & 57.91002 & 1339-1-6-61-307 & 19.90 & 0.11 & 1.18 & RR Lyrae & P01 &  -0.582 & -0.106 & ---\\
260.42271 & 57.87652 & 1339-1-6-62-301 & 21.52 & 0.36 & 0.26 & Unknown  & P01 &   0.555 &  0.000 & ---\\
260.42448 & 57.90823 & 1339-1-6-61-584 & 20.40 & 0.22 & 1.01 & RR Lyrae & P01 &   0.537 &  0.346 & ---\\
260.44703 & 57.88509 & 1339-1-6-62-293 & 20.24 & 0.21 & 1.14 & RR Lyrae & P01 &   0.302 &  0.084 & ---\\
260.45115 & 57.96817 & 1339-1-6-61-504 & 20.11 & 0.07 & 0.16 & RR Lyrae & P01 &  -0.538 & -0.326 & QSO\\
\enddata

\tablenotetext{\dag}{BS61,S79,CS86,P01,G98} \tablenotetext{\ddag}{Baade \& Swope Variables found in BS61 Table B}
\tablenotetext{*}{Not variable in this catalog}
\tablenotetext{1}{CS86 Field 1} \tablenotetext{2}{CS86 Field 2}
\tablenotetext{c}{P01 data from Field C0 unless otherwise noted} \tablenotetext{e}{P01 Field E1}

\end{deluxetable}

\clearpage

\setcounter{page}{0}

\plotone{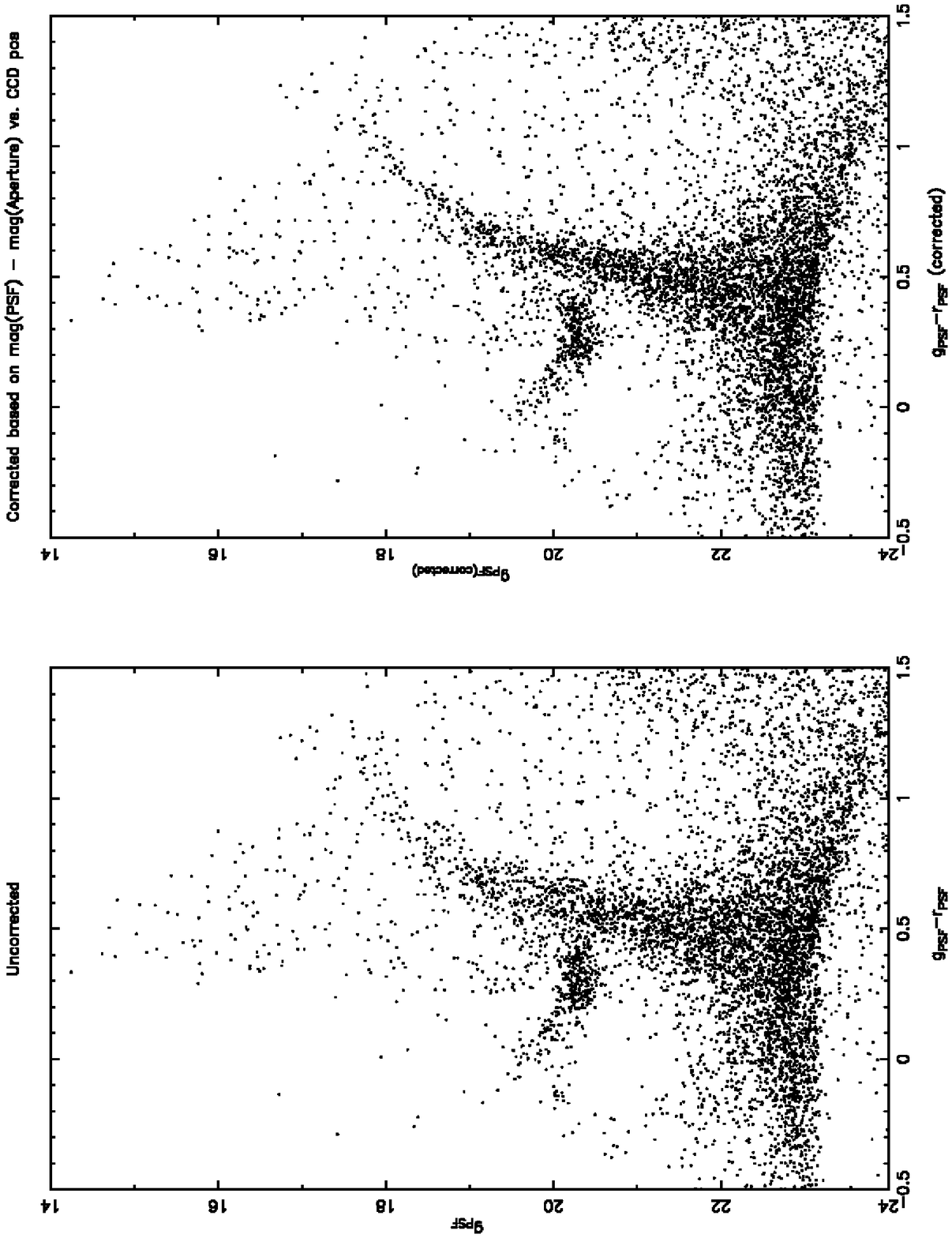}

\plotone{f2.eps}

\plotone{f3.eps}

\plotone{f4.eps}

\plotone{f5.eps}

\plotone{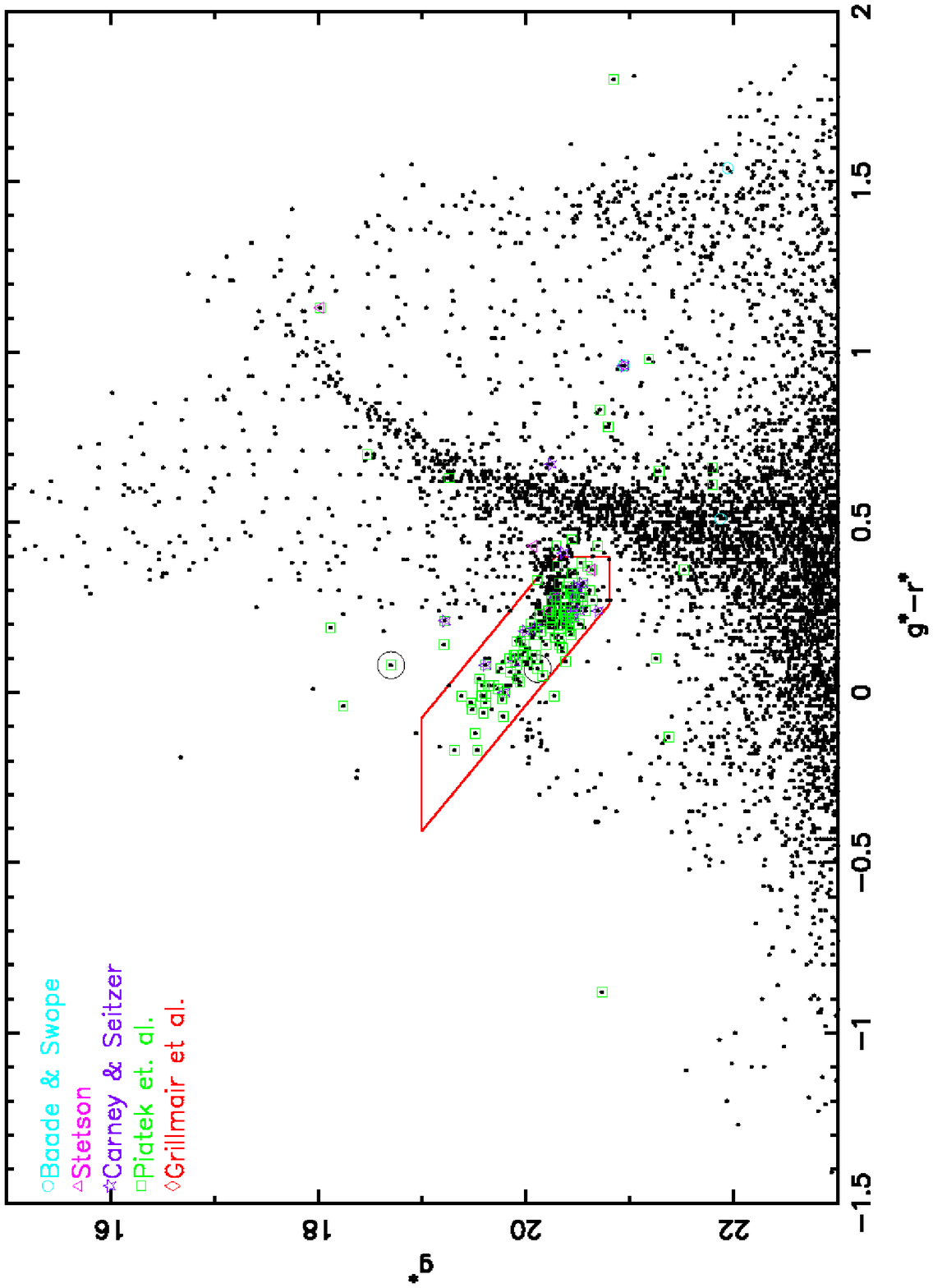}

\end{document}